\documentclass[12pt]{article}

\pdfoutput=1

\usepackage{graphics}
\usepackage{epsfig}
\usepackage{amsfonts}
\usepackage{amssymb}
\usepackage{amsmath}
\usepackage{dsfont}
\usepackage{pifont}
\usepackage{bbm}
\usepackage{latexsym}
\usepackage{verbatim}
\usepackage{cite}
\usepackage{cancel}
\usepackage{slashed}
\usepackage{url}
\usepackage{hyperref}

\newcommand{\be}{\begin{equation}}
\newcommand{\ee}{\end{equation}}
\newcommand{\ben}{\begin{displaymath}}
\newcommand{\een}{\end{displaymath}}
\newcommand{\bea}{\begin{eqnarray}}
\newcommand{\eea}{\end{eqnarray}}

\newcommand{\bean}{\begin{eqnarray*}}
\newcommand{\eean}{\end{eqnarray*}}

\DeclareMathAlphabet{\mathpzc}{OT1}{pzc}{m}{it}

\begin{document}
\pagestyle{plain}


\makeatletter \@addtoreset{equation}{section} \makeatother
\renewcommand{\thesection}{\arabic{section}}
\renewcommand{\theequation}{\arabic{equation}}
\renewcommand{\thefootnote}{\arabic{footnote}}


\setcounter{page}{1} \setcounter{footnote}{0}


\begin{titlepage}

\begin{flushright}
ROM2F/2018/06\\
UUITP-50/18\\
\end{flushright}

\bigskip

\begin{center}

\vskip 0cm

{\LARGE \bf A new light on the darkest corner\\[2mm] of the landscape} \\[6mm]

\vskip 0.5cm



\vspace{1 cm} {\large Johan Bl{\aa}b{\"a}ck$^\star$, Ulf Danielsson$^\dagger$ and Giuseppe Dibitetto$^\dagger$}\\

\vspace{0.85 cm}
{$^\star$Dipartimento di Fisica \& Sezione INFN\\ Università di Roma ``Tor Vergata''\\
Via della Ricerca Scientifica 1, 00133 Roma, Italy}\\
\vspace{0.5 cm}
{$^\dagger$Institutionen f\"or fysik och astronomi, \\
Uppsala Universitet, \\
Box 803, SE-751 08 Uppsala, Sweden}

\vspace{0.7cm}

{\small\upshape\ttfamily $^\star$johan.blaback @roma2.infn.it,} \\
{\small\upshape\ttfamily $^\dagger$\{ulf.danielsson, giuseppe.dibitetto\} @physics.uu.se} \\


\vskip 0.8cm

\end{center}

\vskip 1cm

\begin{center}

{\bf ABSTRACT}\\[3ex]

\begin{minipage}{13cm}
\small

Motivated by the recently proposed bounds on the slow-roll parameters for scalar potentials arising from string/M-theory compactifications, a.k.a.~the Refined de Sitter Swampland conjecture,
we explore the sharpness of such constraints within 4D supergravities coming from compactifications of massive type IIA string theory on $\mathbb{T}^{6}/(\mathbb{Z}_{2}\,\times\,\mathbb{Z}_{2})$.
With the aid of a numerical technique, known as differential evolution, we are able to find a de Sitter extremum which is fully metastable up to one single flat direction.
This solution is supported by spacetime filling sources such as O6 planes and KK monopoles.
Our example violates the bound imposed by this conjecture.

\end{minipage}

\end{center}

\vfill

\end{titlepage}



\section*{Introduction}

Just before the turn of the millennium, measurements of the expansion of the universe using supernovae led to the discovery of dark energy \cite{Riess:1998cb,Perlmutter:1998np}. This combined with measurements of the Cosmic Microwave Background (CMB) radiation \cite{Jaffe:2000tx} and the Baryonic Acoustic Oscillations (BAO) \cite{Tegmark:2003ud}, gave rise to the currently preferred $\Lambda$CDM model of cosmology, according to which our universe has a positive and small cosmological constant.

As a consequence, many research efforts in the last two decades have been focused on finding metastable de Sitter (dS) solutions within string theory. Despite the initial enthusiasm, it has been remarkably difficult to construct rigorous string theory models admitting dS vacua within a reliable regime. All known constructions make use of ingredients that are not well under control. This includes KKLT \cite{Kachru:2003aw}, where non-perturbative effects play a crucial role. See \cite{Danielsson:2018ztv} for a review of the difficulties with these and other attempts.

A particularly interesting set of models are the massive type IIA compactifications, which were among the first examples where complete moduli stabilization could be established \cite{DeWolfe:2005uu}.
This result was achieved through a Calabi-Yau compactification with both NS-NS and R-R fluxes, as well as O6 planes. Moreover, parameters could be tuned so that the resulting AdS vacuum ends up in a perturbative regime with large volume and small string coupling. Extending this to dS solutions, the no-go theorem of \cite{Hertzberg:2007wc} applies, resulting in a lower bound of $\mathcal{O}(1)$ for the first slow-roll parameter, thus excluding the existence of dS vacua with a flat internal manifold.

The simplest way out, is to allow for negative curvature in the compact dimensions. Substantial progress relevant to these IIA constructions has been achieved through the use of the associated minimal supergravity description in four dimensions (see \emph{e.g.} \cite{Giddings:2001yu,Kachru:2002he, Derendinger:2004jn, DeWolfe:2004ns,
Camara:2005dc, Villadoro:2005cu, Derendinger:2005ph, Aldazabal:2006up, Aldazabal:2007sn,DallAgata:2009wsi}). Still, there were many hints suggesting that fully metastable dS vacua, as well as quintessence, could be ruled out.

Within the isotropic sector of the model, the best one can find is a family of tachyonic dS solutions with second slow-roll parameter of $\mathcal{O}(-1)$ \cite{Caviezel:2008tf, Danielsson:2009ff, deCarlos:2009fq, Danielsson:2010bc, Dibitetto:2010rg, Danielsson:2011au,
Danielsson:2012et,Junghans:2016uvg,Junghans:2016abx}. Later in \cite{Blaback:2013fca}, by exploring the non-isotropic sector of the model, a few examples of quintessence solutions were found. These configurations display a running dark energy characterized by both slow-roll parameters of $\mathcal{O}(10^{-1})$, resulting in a few e-folds of cosmic acceleration. The tachyonic dS solutions serve as counterexamples to the swampland conjecture of \cite{Obied:2018sgi} -- at least if issues of quantization of fluxes are ignored (see \cite{Roupec:2018mbn}). They do satisfy, however, the very recent refined de Sitter swampland conjecture of \cite{Ooguri:2018wrx} (in line with the original proposals of \cite{Andriot:2018wzk,Garg:2018reu,Garg:2018zdg}).

If non-geometric fluxes are added to the massive IIA theory compactified on twisted tori, it is easy to find metastable dS \cite{Font:2008vd,deCarlos:2009qm,Danielsson:2012by,Blaback:2013ht} that does violate the constraints set by that conjecture. The status of these models is, however, not clear. As explored in \cite{Danielsson:2015tsa,Blaback:2015zra} some combinations of non-geometric fluxes might be locally geometric and just correspond to non-trivial topologies. So far, there do not exist any such examples yielding metastable dS.

In the present paper we return to the non-isotropic sectors of the $\mathcal{N}=1$ supergravity theories arising from massive type IIA compactifications with O$6$/D$6$ on $\mathbb{T}^{6}/(\mathbb{Z}_{2}\,\times\,\mathbb{Z}_{2})$, which were found to contain quintessence-like configurations \cite{Blaback:2013fca}. In that work, the metric fluxes obeyed the Jacobi constraint thus ruling out the presence of KK monpoles\footnote{In \cite{Derendinger:2014wwa} these sources were employed to discuss the possibility of new $\textrm{AdS}_4$ solutions in the context of M-theory compactifications on G$_2$ structure manifolds.} \cite{Villadoro:2007yq}. As we will show in this paper, something quite remarkable happens when this constraint is relaxed. In the presence of spacetime filling KK monopoles, the otherwise persistent tachyonic direction can be made to disappear. All directions have now positive squared masses, except one that is flat.

The flat direction of the potential is axionic and always present if the superpotential is at most linear in the $S$ \& $T$ moduli. It is present in earlier searches for dS critical points, but always overshadowed by the notorious tachyons. In this paper we show that the tachyons can be removed through the use of KK monopoles\footnote{It is interesting to note that the possible importance of KK monopoles in obtaining metastable dS was already suggested in \cite{Silverstein:2007ac}.}, while a strictly metastable dS would require the addition of terms like $ST$ or $T^{2}$, \emph{i.e.}, non-geometric terms.

We note that the classical flat direction of our solution could in principle receive corrections from quantum or non-pertubative effects. In this way, slow-roll inflation or quintessence-like solutions might be generated.
It would be interesting to understand whether the universal character of our flat direction has any implications for the objections against quintessence.

\section*{Massive type IIA compactifications with sources}

Reductions of massive type IIA string theory on a twisted $\mathbb{T}^{6}$ with fluxes and O$6$ planes have been extensively studied in the literature.
The presence of orientifold planes as local sources threading internal space enforces a discrete $\mathbb{Z}_{2}^{3}$ reflection symmetry, which turns out
to break supersymmetry down to $1/8$.
The corresponding truncation to parity preserving fields within the compactification process yields a resulting $\mathcal{N}=1$ supergravity model in four dimensions.

In particular, models of this type enjoy $\textrm{SL}(2,\mathbb{R})^{7}$ global bosonic symmetry, arising from the coupling of gravity to seven chiral multiplets
carrying information concerning the parity even closed string sector. The scalar sector contains seven complex fields $\Phi^{\alpha}\,\equiv\,\left(S,T_{i},U_{i}\right)$ with $i=1,2,3$.
The kinetic Lagrangian follows from the K\"ahler potential
\be
\label{Kaehler_STU}
\mathcal{K}\,=\,-\log\left(-i\,(S-\overline{S})\right)\,-\,\sum_{i=1}^{3}{\log\left(-i\,(T_{i}-\overline{T}_{i})\right)}\,-\,\sum_{i=1}^{3}{\log\left(-i\,(U_{i}-\overline{U}_{i})\right)}\ ,
\ee
through $\mathcal{L}_{\textrm{kin}}=\left(\partial_{\alpha}\partial_{\overline{\beta}}\mathcal{K}\right)\,\partial \Phi^{\alpha}\partial\overline{\Phi}^{\overline{\beta}}$.
The presence of non-trivial fluxes and internal curvature induces a scalar potential $V$ for the scalar fields, which is given in terms of the above K\"ahler potential and a holomorphic superpotential $\mathcal{W}$ by
\be
\label{V_N=1}
V\,=\,e^{\mathcal{K}}\left(-3\,|\mathcal{W}|^{2}\,+\,\mathcal{K}^{\alpha\bar{\beta}}\,D_{\alpha}\mathcal{W}\,D_{\bar{\beta}}\overline{\mathcal{W}}\right)\ ,
\ee
where $\mathcal{K}^{\alpha\bar{\beta}}$ is the inverse K\"ahler metric and $D_{\alpha}$ denotes the K\"ahler-covariant derivative.

The general form of a superpotential induced by geometric fluxes in type IIA with O$6$ planes and KK monopoles is given by \cite{Villadoro:2005cu}
\be
\label{W_Geom}
\mathcal{W}\,=\,P_{1}(U_{i})\,+\,S\,P_{2}(U_{i}) \,+\,\sum\limits_{k}{T_{k}\,P_{3}^{(k)}(U_{i})}\ ,
\ee
where $P_{1}$, $P_{2}$ and $P_{3}^{(k)}$ are polynomials in the complex structure moduli given by
\be
\begin{array}{cclc}
P_{1}(U_{i}) & = & a_{0}\,-\,\sum\limits_{i}{a_{1}^{(i)}\,U_{i}}\,+\,\sum\limits_{i}{a_{2}^{(i)}\,\dfrac{U_{1}\,U_{2}\,U_{3}}{U_{i}}}\,-\,a_{3}\,U_{1}\,U_{2}\,U_{3} & , \\[3mm]
P_{2}(U_{i}) & = & -b_{0}\,+\,\sum\limits_{i}{b_{1}^{(i)}\,U_{i}} & , \\[3mm]
P_{3}^{(k)}(U_{i}) & = & c_{0}^{(k)}\,+\,\sum\limits_{i}{c_{1}^{(ik)}\,U_{i}} & .
\end{array}
\ee
The higher dimensional interpretation of the above superpotential couplings as massive type IIA fluxes is explicitly spelled out in Table~\ref{table:fluxes}.

%
%
\begin{table}[h!]
\renewcommand{\arraystretch}{1.25}
\begin{center}
\scalebox{0.92}[0.92]{
\begin{tabular}{ | c || c | c | c |}
\hline
couplings & Type IIA & fluxes & dof's\\
\hline
\hline
$1 $ &  $F_{ambncp}$ & $  a_0 $ & $1$\\
\hline
$U_{i} $ &  $F_{ambn}$ & $   -a_1^{(i)} $ & $3$\\
\hline
$U_{j}U_{k} $ &  $F_{am}$ & $  a_2^{(i)} $ & $3$\\
\hline
$U_{i}U_{j}U_{k} $ & $F_{0}$ & $  -a_3 $ & $1$\\
\hline
\hline
$S $& $ {H}_{mnp} $ & $  -b_0$ & $1$\\
\hline
$S \, U_{i} $ &  ${{\omega}_{mn}}^{c}$ & $  b_1^{(i)} $ & $3$\\
\hline
\hline
$T_{i} $& $ H_{a b p} $ & $  c_0^{(i)} $ & $3$\\
\hline
$T_{i} \, U_{j} $ &  $ {\omega_{p a}}^{n} = {\omega_{b p}}^{m} \,\,\,,\,\,\, {\omega_{b c}}^a $  & $c_1^{(ji)} $ & $9$\\
\hline
\end{tabular}
}
\end{center}
\caption{{\it The relation between type IIA fluxes and superpotential couplings. The six internal directions of $\mathbb{T}^{6}$ are split into $\,``-"$ labelled by $a=1,2,3$, and
$\,``\,|\,"$ labelled by $m=4,5,6$ (\emph{i.e.} parallel and transverse to $\textrm{O}6^{||}$ respectively).
Note that the orbifold involution forces $i,j,k$ to be all different any time they appear as indices of fields of the same type ($T$ or $U$).}}
\label{table:fluxes}
\end{table}
\begin{table}[h!]
\renewcommand{\arraystretch}{1.25}
\begin{center}
\scalebox{0.92}[0.92]{
\begin{tabular}{ | c || c | c | c |}
\hline
sources & tadpoles  & cycle & dof's\\
\hline
\hline
(O6/D6)$^{||}$ &  $a_3b_0\,-\,\sum\limits_{i}a_2^{(i)}b_1^{(i)}$ &  $abc$ & $1$\\
\hline
(O6/D6)$^{\perp}$ &  $a_3c_0^{(i)}\,+\,\sum\limits_{j}a_2^{(j)}c_1^{(ji)}$ &  $anp$ & $3$\\
\hline
\hline
(KK5/KKO5) &  $b_1^{(j)}c_1^{(ji)}\,+\,b_1^{(i)}c_1^{(jj)}$ &  $an;p_{\textrm{ISO}}$ & $6$\\
\hline
(KK5/KKO5)$^{\prime}$ & $c_1^{(ij)}c_1^{(jk)}\,+\,c_1^{(ik)}c_1^{(jj)}$ &  $an;c_{\textrm{ISO}}$ & $6$\\
\hline
\end{tabular}
}
\end{center}
\caption{{\it The relation between type IIA flux tadpoles and the corresponding sources, which entirely fill spacetime, and in addition wrap the specified internal cycle.
Note that the orbifold involution forces $i,j,k$ to be all different any time they appear in a flux tadpole, while repeated indices are not summed over unless
explicitly indicated.}}
\label{table:sources}
\end{table}
%
In terms of the fourteen real fields appearing in the explicit parametrization
\be
\left\{\begin{array}{lclc}
S & = & \chi \, + \, i \, e^{-\phi} & , \\[1mm]
T_{i} & = & \chi^{(1)}_{i} \, + \, i \, e^{-\phi^{(1)}_{i}} & , \\[1mm]
U_{i} & = & \chi^{(2)}_{i} \, + \, i \, e^{-\phi^{(2)}_{i}} & ,
\end{array}\right.
\ee
where $\left\{\phi^{I}\right\} \, \equiv \, \left\{\phi, \, \phi^{(1)}_{i}, \, \phi^{(2)}_{i}, \, \chi, \, \chi^{(1)}_{i}, \, \chi^{(2)}_{i}\right\}$, with $i=1,2,3$,
the effective 4D Lagrangian reads
\be
\label{L1}
\mathcal{L}_{\textrm{eff}} \, = \, \sqrt{-g} \, \left(-\frac{1}{2} \, \mathcal{K}_{IJ}(\phi) \, \partial_{\mu}\phi^{I} \, \partial^{\mu}\phi^{J} \, - \, V(\phi)\right) \ ,
\ee
where $I=1, \cdots, \, 14$ and $\mathcal{K}_{IJ}$ is derived from the K\"ahler metric $\mathcal{K}_{\alpha\bar{\beta}}$ when rewritten in terms of the above real dof's.
Now the physical slow-roll parameters quantitatively describing the flatness properties of the scalar potential $V$ may be defined as
\be
\label{slow-roll_V}
\begin{array}{lclclclc}
\epsilon_{V} & \equiv & \frac{1}{2} \,\mathcal{K}^{IJ} \, \frac{D_{I}V \, D_{J}V}{V^{2}}  & \textrm{ and } & \eta_{V} & \equiv &
\textrm{Min.~Eig.}\left(\frac{\mathcal{K}^{JK} \, D_{I} \, D_{K}V}{|V|}\right)  & .
\end{array}
\ee
Generic expectations based on the recently proposed swampland conjectures would suggest some difficulties in achieving small values for both $\epsilon_{V}$ and
$|\eta_{V}|$ whenever $V>0$, or in terms of \cite{Ooguri:2018wrx}, having $c = \sqrt{2 \epsilon_V}$ and $c' = -\eta_V$ small. In what follows we will illustrate an explicit example of a massive type IIA string theory background admitting a dS extremum where $\epsilon_V=\eta_V=0$.

\section*{The explicit flat de Sitter solution}

There are a multitude of approaches available to solve optimization problems like the one we have in front of us. The approach we use here is similar to that of \cite{Blaback:2013fca}, were solutions are found by tuning the flux values while keeping the moduli fixed at the origin. We are interested in solutions that satisfy the following conditions
\begin{equation}
  V > 0\,,\quad \epsilon_V \ll 1\,,\quad |\eta_V| \ll 1\,.
\end{equation}
Technically this can be implemented in several ways, we chose to optimize the square weighted sum
\begin{equation}
  \alpha_1 (V_\theta)^2 + \alpha_2 (\sqrt{2\epsilon_V})^2 + \alpha_3 (\eta_V)^2\,,
\end{equation}
where $V_\theta$ is a step function which is zero when the potential is positive and is equal to the potential when it is negative. The weights $\alpha_i$ are chosen to prioritize the positivity of the potential, \emph{e.g.}  $\{\alpha_i\} = \{100,1,1\}$, and then proceed to optimize the $\epsilon_V$ and $|\eta_V|$.

To solve this optimization problem we used \texttt{BlackBoxOptim.jl} \cite{Feldt2018} which is an optimization package for the Julia programming language \cite{DBLP:journals/corr/BezansonEKS14}. Using its default algorithm, which is a differential evolution algorithm, we are able to find a solution which has the smallest eigenvalue zero, $\eta_V = 0$, beyond the numerical precision used, and the equations of motion are also solved, that is $\epsilon_V = 0$, numerically. The physical properties of the solutions are displayed in Table~\ref{table:output}, while the explicit solution is given by the flux values displayed in Table~\ref{table:fluxvalues} in Appendix~\ref{app:fluxes}.
Further information concerning the mass spectrum is collected in Table~\ref{table:masses} in Appendix~\ref{app:fluxes}.
\begin{table}[h!]
\begin{center}
\begin{tabular}{|c||c|}
\hline
$V$       & $909.5199533494178$\\
\hline
$\sqrt{2\epsilon_V}$ & $8.300748392626971\times 10^{-15}$\\
\hline
$\eta_V$ & $-1.5148183731394715\times 10^{-26}$\\
\hline
\end{tabular}
\end{center}
\caption{{\it The physical properties of the solution.}}
\label{table:output}
\end{table}

As we can see, this solution does hence not obey the constraints posed by the Refined de Sitter Swampland conjecture, and is the best candidate for a proper counterexample available, in terms of the control we have over the ingredients. It is a novel example of classical de Sitter solution without a tachyon.
Compared to \cite{Blaback:2013fca}, the present search did not enforce the Jacobi constraints on metric flux parameters, \emph{i.e.}
\begin{equation}
  \omega_{[AB}^{\phantom{[AB}E}\omega_{C]E}^{\phantom{C]E}D} \, \neq \, 0\ ,
\end{equation}
which implies the presence of KK monopoles. However, by virtue of the analysis in \cite{Danielsson:2014ria}, there exists an alternative geometric description of these in terms of more general $\textrm{SU}(3)$ structures beyond group manifolds. The corresponding $\omega$ components can be mapped into torsion classes $W_1$, $W_2$ and $W_3$ parametrizing the internal curvature. The numerical values for these are computed in our solution and given in Table~\ref{tab:torsion}. We can also verify using the present algorithm that we get solutions comparable to those of \cite{Blaback:2013fca} once the Jacobi constraints are indeed imposed.
\begin{table}
  \begin{center}
    \begin{tabular}{|c||ccc|}
      \hline
      $W_1$ & & $224.68135481021886$ &\\
      \hline
      $W_2^{(i)}$ & $32.2085108242145$ & $-113.04192473425071$ & $80.83341391003621$ \\
      \hline
      $W_3^{(0)}$ & & $181.89953007912084$ & \\
      \hline
      $W_3^{(i)}$ & $-250.44238257719837$ & $531.0656092543013$ & $-98.72369659798198$ \\
      \hline
    \end{tabular}
    \caption{{\it Numerical values for the torsion classes in our dS solution. These correspond to a negatively curved half-flat SU(3) structure. 
    Note that the primitivity constraints $W_2\wedge J\wedge J = W_3\wedge\Omega = 0$ are satisfied in the origin of moduli space thanks to $\sum\limits_i W_2^{(i)} = 0$ \& $\sum\limits_i W_3^{(i)} = W_3^{(0)}$. \label{tab:torsion}}}
  \end{center}
\end{table}

\subsection*{Potential profile and plots}

\begin{figure}
  \begin{center}
    \begin{tabular}{cc}
    \includegraphics[scale=0.5]{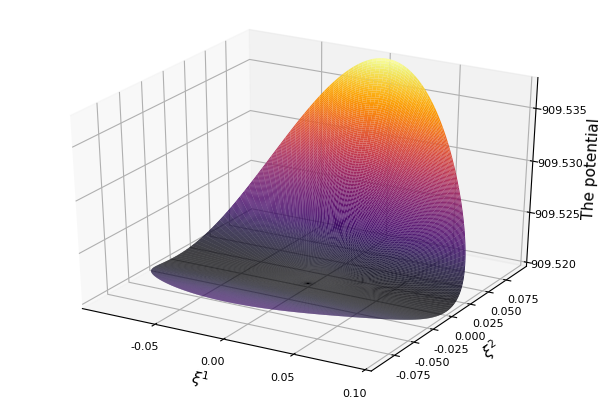}
    &
    \includegraphics[scale=0.5]{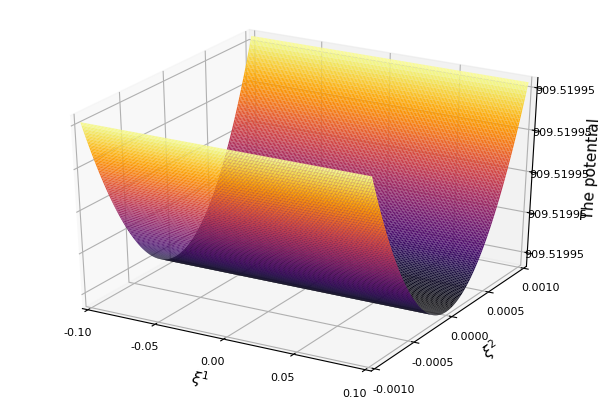}
    \end{tabular}
    \caption{{\it Behavior of the potential along the smallest and next to smallest eigenvalue. To the left the potential is plotted equally far into the moduli space for both directions, while the right picture is a zoom in for the displacement along $\xi^2$, at which the $\xi^1$ still appears flat.} \label{fig:flatplot}}
  \end{center}
\end{figure}

The potential at fixed fluxes, as we have here, is parametrized by the fourteen real moduli, for which our solution is positioned at the origin. Labeling the eigenvalues as $\xi^i$ with $i$ corresponding to increasing mass, \emph{e.g.} $\xi^1$ is the eigenvector corresponding to the smallest eigenvalue, we can plot the potential along directions of two eigenvectors. For instance, the eigendirection $\xi^1$ is, as already mentioned in the introduction, along a linear combination of the axions of $S$ and $T_i$
\begin{equation}
  \xi^1 = 0.443498\,\partial_{\chi} - 0.886153\,\partial_{\chi^{(1)}_1} - 0.0269893\,\partial_{\chi^{(1)}_2} + 0.131581\,\partial_{\chi^{(1)}_3}\ ,
\end{equation}
up to numerical zeros.
Note that the presence of such a flat direction within the axionic sector of the spectrum is a universal feature of all models characterized by a superpotential linear in more than one complex field (in this case $S$ \& the three $T_i$'s).
In this situation one can prove that the on-shell potential is completely independent of the aforementioned combination of the axions.
To visually illustrate the flat direction we plotted in Figure~\ref{fig:flatplot} the profile of the potential along $\xi^1$ and $\xi^2$. To show that this does indeed appear flat we have also made a plot that extends less in the massive $\xi^2$ direction to the right in the same figure.

\begin{figure}
  \begin{center}
    \begin{tabular}{cc}
      \includegraphics[scale=0.5]{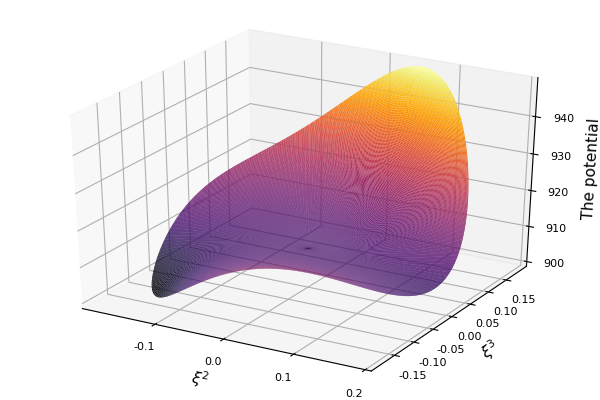}
      &
      \includegraphics[scale=0.5]{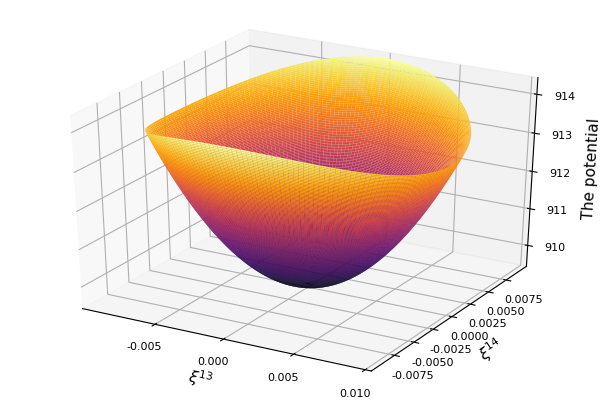}
    \end{tabular}
    \caption{{\it A plot of the potential along the eigenvectors corresponding to positive eigenvalues. To the left we see displacements along the eigenvectors corresponding to the first two non-zero eigenvalues. To the right shows the potential along the eigenvectors corresponding to the two highest masses.}\label{fig:23plot}}
  \end{center}
\end{figure}

The second smallest eigenvalue is still fairly small, $0.00383681$, but not close to a numerical zero. When the potential is plotted along $\xi^2$ and $\xi^3$, see left plot in Figure~\ref{fig:23plot}, we can see that indeed that mass is small enough to be invisible to the naked eye, and that sufficiently far away from the origin the potential dips down. For the two most massive directions, $\xi^{13}$ and $\xi^{14}$, we can clearly see the potential being stable, as illustrated in the right plot in Figure~\ref{fig:23plot}.

\subsection*{Summary}

We have found a non-tachyonic dS solution in massive type IIA with only geometric fluxes. All directions are stabilized except one that is flat. It is crucial for our construction to break the Jacobi constraint on 
the metric fluxes and thus introduce KK monopoles. By mapping the fluxes to $\textrm{SU}(3)$ structured torsion forms, the KK monopoles are seen to be nothing else than aspects of pure geometry. If our solution turned out to be a counterexample to the refined swampland conjecture, it would be interesting to see whether it can be developed into a fully realistic cosmology.

%
%

\section*{Acknowledgements}

We would like to thank Thomas Van Riet and Raffaele Savelli for interesting and stimulating discussions.
The work of JB is supported by the MIUR-PRIN contract 2015MP2CX4002 {\it ``Non-perturbative aspects of gauge theories and strings''}. The work of UD \& GD is supported by the Swedish Research Council (VR).

\appendix

\section{Explicit values for the solution}
\label{app:fluxes}

\begin{table}[h!]
\begin{center}
\begin{tabular}{|c||c|}
\hline
& Solution\\
\hline
$a_0$       & $-217.76872424686914$\\
\hline
$a_1^{(i)}$ & $\begin{array}{ccc}-41.79763020180156&70.56487102375863&-89.22430722425253\end{array}$\\
\hline
$a_2^{(i)}$ & $\begin{array}{ccc}-23.33306230714147&189.9460634547514&-154.0049022373281\end{array}$\\
\hline
$a_3$       & $-464.5883839179391$\\
\hline
$b_0$       & $100.08413177965302$\\
\hline
$b_1^{(i)}$ & $\begin{array}{ccc}-426.43001031208365&349.5859385925167&231.96657385577444\end{array}$\\
\hline
$c_0^{(i)}$ & $\begin{array}{ccc}-82.45853019965551&57.84106986098164&-206.13035263716245\end{array}$\\
\hline
$c_1^{(ij)}$& $\begin{array}{ccc}-162.6594420362874&417.686913244789&427.51418377873284\\209.00223264058445&281.9279790429654&287.09810774987557\\40.23685903383291&168.47274918187517&-476.3139559112621\end{array}$\\
\hline
\end{tabular}
\end{center}
\caption{{\it The flux values for our de Sitter solution.}}
\label{table:fluxvalues}
\end{table}
\begin{table}[h!]
\begin{center}
\begin{tabular}{|c c c c c|}
\hline
& & Mass Spectrum & & \\
\hline
210.788 & & 168.559 & & 144.436\\
98.268 & & 75.190 & & 69.841\\
41.773 & & 33.389 & & 25.499\\
19.899 & & 4.700 & & 1.968\\
0.004  & & 0 &  & \\
\hline
\end{tabular}
\end{center}
\caption{{\it The physical mass spectrum for our de Sitter solution. The numerical values are calculated as the eigenvalues of $\frac{\mathcal{K}^{JK} \, D_{I} \, D_{K}V}{|V|}|_{\mathrm{sol.}}$, and truncated to a precision of $10^{-3}$.}}
\label{table:masses}
\end{table}

%
%

\small


\bibliography{references}
\bibliographystyle{utphys}

\end{document}